\pdfoutput=1
\documentclass[%
 reprint,
superscriptaddress,
 amsmath,amssymb,
 aps,
pre,
]{revtex4-1}

\usepackage{graphicx}
\usepackage{dcolumn}
\usepackage{color}	
\usepackage{bm}
\usepackage{braket}

\begin{document}


\title{X-ray Thomson scattering spectra from DFT-MD simulations based on a modified Chihara formula}

\author{Maximilian \surname{Schörner}}
 \email{maximilian.schoerner@uni-rostock.de}
\affiliation{%
 University of Rostock, Institute of Physics, 18051 Rostock, Germany
}%

\author{Mandy \surname{Bethkenhagen}}
\affiliation{%
 École Normale Supérieure de Lyon, Laboratoire de Géologie de Lyon LGLTPE UMR 5276, Centre Blaise Pascal, 46 allée d’Italie Lyon 69364, France
}%
\affiliation{%
 Institute of Science and Technology Austria, Am Campus 1, 3400 Klosterneuburg, Austria
}%

\author{Tilo \surname{Döppner}}
\affiliation{%
 Lawrence Livermore National Laboratory, Livermore, CA 94551, USA
}%

\author{Dominik \surname{Kraus}}
\affiliation{%
 University of Rostock, Institute of Physics, 18051 Rostock, Germany
}%
\affiliation{%
 Helmholtz-Zentrum Dresden-Rossendorf, 01328 Dresden, Germany
}%

\author{Siegfried H. \surname{Glenzer}}
\affiliation{%
 SLAC National Accelerator Laboratory, Menlo Park, CA 94025, USA
}%

\author{Ronald \surname{Redmer}}
\affiliation{%
 University of Rostock, Institute of Physics, 18051 Rostock, Germany
}%





\date{\today}

\begin{abstract}
We study state-of-the-art approaches for calculating x-ray Thomson scattering spectra from density functional theory molecular dynamics (DFT-MD) simulations based on a modified Chihara formula that expresses the inelastic contribution in terms of the dielectric function. We compare the electronic dynamic structure factor computed from the Mermin dielectric function using an \textit{ab~initio} electron-ion collision frequency to computations using a linear response time dependent density functional theory (LR-TDDFT) framework for hydrogen and beryllium and investigate the dispersion of free-free and bound-free contributions to the scattering signal.
A separate treatment of these contributions in the Mermin dielectric function shows excellent agreement with LR-TDDFT results for ambient-density beryllium, but breaks down for highly compressed matter where the bound states become pressure ionized.
LR-TDDFT is used to reanalyze x-ray Thomson scattering experiments on beryllium demonstrating strong deviations from the plasma conditions inferred with traditional analytic models at small scattering angles.
\end{abstract}

\maketitle


\section{Introduction}
\label{sec:intro}
X-ray Thomson scattering (XRTS) has been one of the premier diagnostic tools for warm dense matter (WDM) experiments, enabling measurements of the electron density, temperature and ionization state~\cite{Glenzer2009,Fletcher2015,Faeustlin2010}. The states reached in these experiments are characterized by temperatures of a few electronvolts (eV) and around solid densities, which constitutes strongly correlated plasmas with non-negligible degeneracy. This prevents the application of ideal plasma theory for the analysis of these experiments, and rather requires a quantum mechanical treatment in a many-body framework.
Knowledge of equation of state data as well as thermal and electrical transport properties for warm dense hydrogen and beryllium is essential for modelling astrophysical objects~\cite{Guillot2005,Helled2011} and inertial confinement fusion~\cite{Lindl2004}, where hydrogen is used as fuel while beryllium often serves as ablator material~\cite{Xu2007,Simakov2014}. Furthermore, hydrogen and beryllium are excellent test cases for new theoretical approaches. The analytical behavior in many limiting cases for fully ionized hydrogen plasmas are known and beryllium can be used to test the treatment of bound states in a simple low-$Z$ material.
WDM is typically opaque in the optical regime, as the light frequency is smaller than the plasma frequency $\omega_\mathrm{pl}$ of these plasmas. Therefore, it is indispensable to have diagnostic tools at experiments that are well understood, both experimentally and theoretically. XRTS has proven to overcome many of the experimental challenges of probing WDM. The high energy x-ray photons can penetrate dense plasmas and since the advent of free electron lasers (FEL), rep-rated x-ray sources with sufficient brilliance for probing short-lived transient states are available in addition to laser-plasma sources which only allow a limited number of experiments and require complex sample assemblies. New FEL techniques like self-seeding~\cite{Seeding2012, McBride2018} have also resulted in much narrower bandwidths of the x-ray source, enabling the measurement of phonons and ion acoustic modes~\cite{Descamps2020,Wollenweber2021} and a better resolution of density and temperature-sensitive regions in the XRTS spectrum.

Due to the steadily improving quality of collected spectra, it is vital to have accurate theoretical modeling of the scattering. While in the past, the resolution of XRTS spectra often did not allow for discrimination between different theoretical approaches, now, fitting experimental spectra to theoretical models has allowed predictions of electron temperature and density to within a few percent uncertainties~\cite{Frydrych2020,MacDonald2021,Fletcher2022}. As a result, the fidelity of the theoretical model used is now the limiting factor in determining the correct plasma parameters in experiments that employ XRTS as a diagnostic tool. Most approaches rely on the semi-classical Chihara decomposition~\cite{Chihara1987, Chihara2000} of the spectrum into three distinct contributions which originates from distinguishing between free and bound electrons in a chemical picture. An analogous fully quantum mechanical description has also been proposed~\cite{Crowley2014}.
The standard approach for modeling XRTS spectra in the Chihara description is a combination of theories to describe each component individually~\cite{Chapman2014}. The ion dynamics are usually described by the hypernetted-chain approximation with different expressions for the interaction potential while the form factors are described by a screened hydrogenic approximation to the wave functions~\cite{Pauling1932} and the Debye-Hückel approximation for the screening cloud. The plasmon can be described by the random phase approximation (RPA) or the Mermin dielectric function in order to also include electron-ion collisions which can also be approximated to different degrees~\cite{Reinholz2000}. Further electron correlations can be accounted for by local field corrections~\cite{Fortmann2010}. Contributions that are related to bound-free transitions are treated within the impulse approximation~\cite{Eisenberger1970} which is sometimes modified by the ionization potential depression and normalized according to different sum rules. Each of these theories entails various approximations and achieves different degrees of fidelity in a wide range of temperatures and densities which severely complicates gauging sources of errors.

Furthermore, the ionization degree is often left as an unconstrained parameter in the fitting procedure which risks overfitting and neglects its dependence on temperature and density.
In recent years, this approach has been partially replaced by \textit{ab~initio} descriptions like density functional theory molecular dynamics (DFT-MD) simulations and real time or linear response time dependent DFT (RT/LR-TDDFT) computations. Witte~\textit{et~al.} successfully used electron-ion collision frequencies determined by DFT to accurately model the plasmon of an aluminum plasma~\cite{Witte2017}. This approach was subsequently compared to LR-TDDFT by Ramakrishna~\textit{et~al.} for ambient and extreme conditions in aluminum~\cite{Ramakrishna2021} and carbon~\cite{Ramakrishna2019}, which was then used to discern miscibility in an XRTS experiment~\cite{Frydrych2020}. Baczewski~\textit{et~al.} went beyond the Chihara decomposition by simulating the real time propagation of the electronic density using RT-TDDFT~\cite{Baczewski2016}. Path integral Monte Carlo simulations have delivered approximation-free results for the uniform electron gas~\cite{Dornheim2018} and hydrogen plasmas~\cite{Boehme2022}, but are currently unable to describe heavier elements.

In this work, we describe the calculation of XRTS spectra in an \textit{ab initio} framework where each contribution to the Chihara decomposition is extracted from a DFT-MD simulation that is based on well defined approximations and limits the free parameters to the mass density and temperature by constraining the number of free electrons per atom according to a new technique recently proposed by Bethkenhagen \textit{et al.}~\citep{Bethkenhagen2020}. The capability of DFT-MD to compute ion dynamics and the form factors was already demonstrated and tested in previous publications~\cite{Rueter2014, Plagemann2015, Witte2017_ion}. Therefore, we focus on the inelastic contribution that can be computed from a self-consistent DFT cycle using the Kubo-Greenwood formula or LR-TDDFT.
We give an overview of the theoretical foundation for computing the electronic dynamic structure factor from the Mermin dielectric function with a dynamic complex collision frequency and apply this framework to extract a DFT-based collision frequency in Secs.~\ref{sec:structurefactor} and \ref{sec:diel}. In Sec.~\ref{sec:DFTMD} we give the details of the simulation method. We compute DFT-based collision frequencies for a hydrogen plasma and compare them to several analytic approaches in Sec.~\ref{sec:dynColl} and we study the impact of these collision frequencies on DSFs for hydrogen and beryllium plasmas in Secs.~\ref{sec:electronDSF-H}, \ref{sec:electronDSF-isoBe}, and \ref{sec:electronDSF-compBe}.
In Sec.~\ref{sec:experiments}, we apply LR-TDDFT to XRTS experiments on beryllium to evaluate their impact on the inferred plasma parameters.

\section{Theoretical background}
\label{sec:theory}
\subsection{Dynamic structure factor}
\label{sec:structurefactor}
The electronic dynamic structure factor (DSF)~\cite{Glenzer2009}
\begin{equation} 
  \label{eq_seetot}
S_{ee}^{\mathrm{tot}}(\vec{k}, \omega) = \frac{1}{2 \pi N_e} \int_{-\infty}^{\infty} \mathrm{d}t \langle n_{\vec{k}}^e (\tau) n_{-\vec{k}}^e (\tau + t) \rangle_\tau \, e^{i \omega t}
\end{equation}
is the central quantity representing the spatially resolved power spectrum of an electronic system, describing its dynamics at given temporal and spatial periodicities given by the frequency $\omega$ and the wave vector $\vec{k}$, respectively. The number of considered electrons is $N_e$ and the spatial Fourier components of the electron density are given by $n_{\vec{k}}^e$. The time is described by $t$ and $\tau$, where $\langle ... \rangle_\tau$ describes a time average over $\tau$. Experimentally, $S_{ee}^{\mathrm{tot}}(\vec{k}, \omega)$ can be used to identify how strong a photon will couple to density fluctuations at a given energy transfer and scattering angle~\cite{Glenzer2009}. In this work we will use a slight modification of the common decomposition of Eq.~\eqref{eq_seetot} introduced by Chihara~\cite{Chihara1987, Chihara2000}:
\begin{multline} 
  \label{eq_chihara}
S_{ee}^{\mathrm{tot}}(\vec{k}, \omega) = \vert f_\mathrm{i}(\vec{k}) + q(\vec{k}) \vert^2 S_\mathrm{ii}(\vec{k},\omega) + \\ + \underbrace{Z_\mathrm{f} S_{ee}^0(\vec{k},\omega) + Z_\mathrm{b} S_\mathrm{bf}(\vec{k}, \omega)}_{Z \, S_\mathrm{et} (\vec{k}, \omega)} .
\end{multline}
The first term refers to the elastic response of the electrons which follow the ion motion described by the ion-ion structure factor $S_\mathrm{ii}(\vec{k},\omega)$. Here, $f_\mathrm{i}(\vec{k})$ describes the contribution of tightly bound electrons and $q(\vec{k})$ represents the loosely bound screening cloud around the ions. The second term, called the electron feature, arises from the collective behavior of the free electrons in the system undergoing transitions to different free-electron states. The number of free electrons per atom is labeled $Z_\mathrm{f}$ and their DSF is denoted by $S_{ee}^0(\vec{k},\omega)$. The last term in Eq.~\eqref{eq_chihara} is the bound-free contribution. In the original work, Chihara clearly separates free and bound electrons and describes this term as a convolution of the DSF of the core electrons with the self-part of the ionic DSF~\cite{Chihara2000}. We treat the bound-free contribution on the same footing as the free electron contibution and introduce the bound-free DSF $S_\mathrm{bf}(\vec{k}, \omega)$ and the number of bound electrons per atom $Z_\mathrm{b}$. Both the free electron and bound-free contributions arise due to inelastic transitions of the electrons and can, therefore, be combined into one DSF $S_\mathrm{et}(\vec{k}, \omega)$ that accounts for all electronic transitions. This avoids the artificial separation into bound and free electrons for both the charge state $Z$ and the DSF. According to the fluctuation-dissipation theorem~\cite{Kubo1966}, this combined DSF can be related the dielectric response described by the dielectric function $\epsilon (\vec{k}, \omega)$ via
\begin{equation} 
  \label{eq_FDT}
S_\mathrm{et}(\vec{k},\omega) = - \frac{\epsilon_0 \hbar \vec{k}^2}{\pi e^2 n_e} \frac{\operatorname{Im}\left[ \epsilon^{-1}(\vec{k}, \omega) \right]}{1-\exp \left(\frac{-\hbar \omega}{k_\mathrm{B} T_e}\right)}.
\end{equation}
The vacuum permittivity is denoted by $\epsilon_0$, the reduced Planck constant is $\hbar$ and $e$ is the elementary charge. The electron density is given by $n_e$, the electron temperature is $T_e$ and the Boltzmann constant is $k_\mathrm{B}$.
At which conditions the separation into free and bound-free part in Eq.~\eqref{eq_chihara} is justified and yields the same results as the combined approach is discussed in Secs.~\ref{sec:electronDSF-isoBe} and \ref{sec:electronDSF-compBe}.

\subsection{Dielectric Function with electron-ion collisions}
\label{sec:diel}
The dielectric function $\epsilon(\vec{k},\omega)$ is a central material property that is connected to other material properties, like the electrical conductivity $\sigma(\omega)$ in the long wavelength limit or the DSF via the fluctuation dissipation theorem from Eq.~\eqref{eq_FDT}. One of the first approaches that produced collective features of the electron system, like plasmons, is the Lindhard dielectric function~\cite{Lindhard1954}
\begin{multline} 
  \label{eq_eps_rpa}
\epsilon^\mathrm{RPA} (\vec{k}, \omega ) = \lim_{\eta \to 0} \bigg[ 1 - \\ - \frac{2e^2}{\epsilon_0 k^2} \, \int \frac{\mathrm{d}^3q}{\left(2\pi\right)^3} \, \frac{f_{\vec{q} - \frac{\vec{k}}{2}} - f_{\vec{q} + \frac{\vec{k}}{2}}}{\hbar \left( \omega + i \eta \right) + E_{\vec{q} - \frac{\vec{k}}{2}} - E_{\vec{q} + \frac{\vec{k}}{2}}  } \bigg].
\end{multline}
which accounts for electric field screening in the Random Phase Approximation (RPA). The arguments $\vec{k}$ and $\omega$ are the wave vector and the angular frequency, respectively while the electron charge is denoted by $e$. $E_{\vec{q}}$ and $f_{\vec{q}}$ are the kinetic energy and the Fermi occupation of an electron with wave vector $\vec{q}$ in the unperturbed free electron gas. The small imaginary contribution to the frequency $\eta$ is introduced to avoid the pole in the integration and approaches zero thereafter. However, for degenerate, strongly correlated systems electron-ion interactions, which are neglected in Eq.~\eqref{eq_eps_rpa}, have to be accounted for in order to accurately describe the dielectric function.

It was shown that electron-ion collisions can be included via a dynamic collision frequency $\nu(\omega)$ in the framework of the Mermin dielectric function~\cite{Mermin1970,Selchow1999,Roepke1999,Millat2003}
\begin{multline} 
  \label{eq_mermin}
\epsilon^\mathrm{Mermin}(\vec{k}, \omega; \nu(\omega)) = 1+ \\ + \frac{\left(1 + i \frac{\nu(\omega)}{\omega} \right) \left( \epsilon^\mathrm{RPA}(\vec{k}, \omega + i \nu(\omega)) - 1 \right)}{1 + i \frac{\nu(\omega)}{\omega} \frac{\epsilon^\mathrm{RPA}(\vec{k}, \omega + i \nu(\omega)) - 1}{\epsilon^\mathrm{RPA}(\vec{k}, 0) -1}}.
\end{multline}
Extensive work has been performed on the evaluation of different analytic collision frequencies and local field corrections~\cite{Reinholz2000,Thiele2008,Fortmann2010}, as well as first attempts to incorporate \textit{ab initio} results to determine collision frequencies~\cite{Plagemann2012}.

We present the derivation of the RPA dielectric function in the presence of a dynamic complex collision frequency in Appendix~\ref{sec:app-rpa}.
Equations~\eqref{eq_mermin}, \eqref{eq_eps_rpa_re3} and \eqref{eq_eps_rpa_im3} are the basis for calculating the Mermin dielectric function for a given dynamic collision frequency $\nu(\omega)$. In the following, because we are dealing with isotropic systems, we will only consider the magnitude of wave vector $\vec{k}$ and drop the vector notation.

One of the most prominent approximations for the collision frequency is the Born collision frequency~\cite{Reinholz2000}, the combination of which with the Mermin dielectric function in Eq.~\eqref{eq_mermin} is called the Born-Mermin approximation~(BMA). It is widely used in the analysis of XRTS spectra in the WDM field. We give the exact equations used in this work in Appendix~\ref{sec:app-born}.
However, complex many-particle effects, as they are considered in \textit{ab~initio} simulations, cannot be accounted for by this approach.
\begin{figure}[htb]
\center{\includegraphics[angle=0,width=1.0\linewidth]{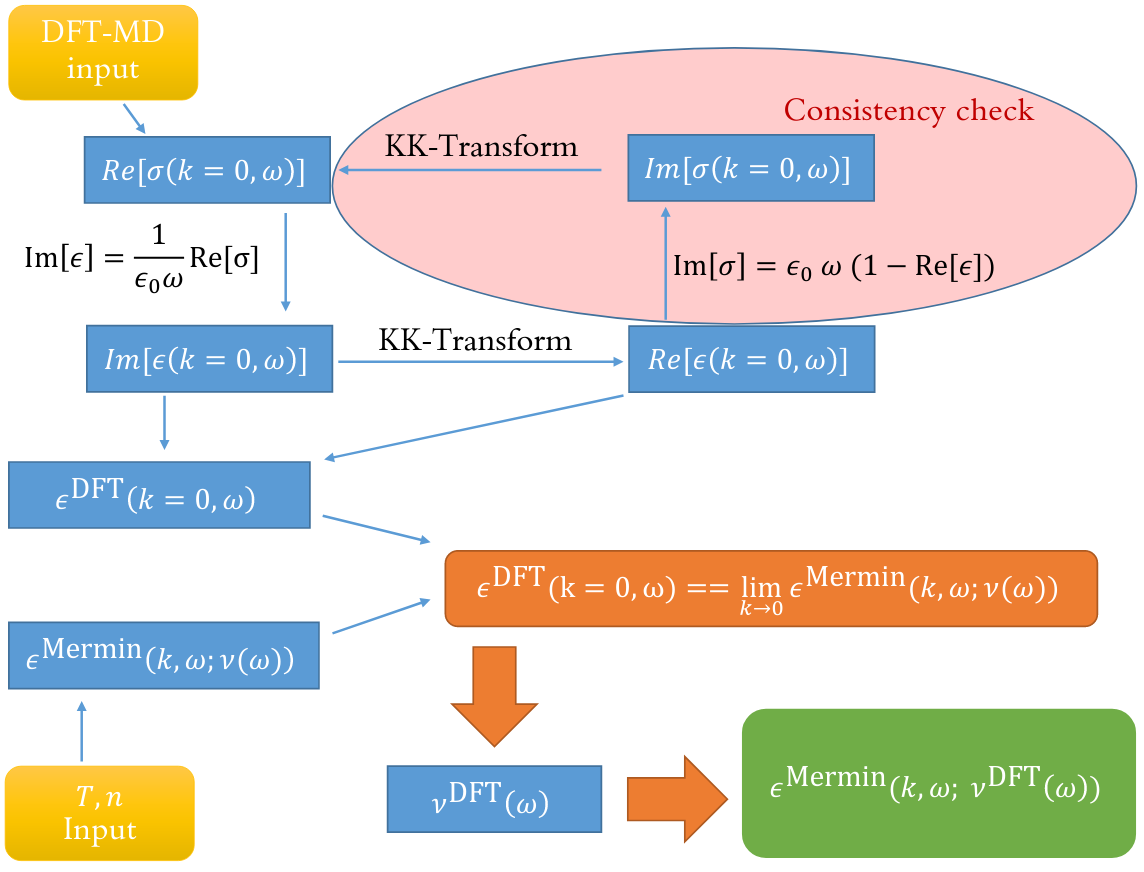}}
\caption{Schematic work flow for determining the dynamic collision frequency and $k$-dependent dielectric function via DFT.}
\label{fig:flowchart}
\end{figure}

In Fig.~\ref{fig:flowchart}, we show the schematic procedure to compute a DFT-based collision frequency from an electrical conductivity in the optical limit. In essence, we construct a complex collision frequency for which the Mermin dielectric function coincides with the \textit{ab~initio} dielectric function in the optical limit. As input, the temperature and electron density of the plasma are needed for the Mermin dielectric function and the real part of the electrical conductivity is needed from the simulation.
According to the Kubo-Greenwood formula~\cite{French2017, Gajdos2006} the conductivity is
\begin{multline} 
  \label{eq_KG}
\operatorname{Re} \left[\sigma (k=0, \omega)\right] = \frac{2 \pi e^2}{3 \omega \Omega} \sum_{\vec{g}} w_{\vec{g}} \sum_{j=1}^N \sum_{i=1}^N  \sum_{\alpha=1}^3 \times \\ \times \big[ f(\epsilon_{j,\vec{g}})- f(\epsilon_{i,\vec{g}}) \big]  |\braket{\psi_{j,\vec{g}}| \hat{v}_\alpha | \psi_{i, \vec{g}}}|^2 \delta(\epsilon_{i,\vec{g}} -\epsilon_{j,\vec{g}} - \hbar \omega ).
\end{multline}
The indices $i$ and $j$ run over the eigenstates, $\alpha$ runs over the spatial orientations and $\vec{g}$ denotes the reciprocal vectors in the Brillouin zone where the wave functions $\psi_{i, \vec{g}}$ are evaluated. The Fermi-Dirac occupation at a given eigenenergy $\epsilon_{j,\vec{g}}$ is described by $f(\epsilon_{j,\vec{g}})$ and $\hat{v}_\alpha$ is the velocity operator in the direction $\alpha$. The normalization volume is denoted by $\Omega$ and $w_{\vec{g}}$ is the weigthing of each $k$-point.
We translate the electrical conductivity to the imaginary dielectric function via
\begin{figure}[b]
\center{\includegraphics[angle=0,width=1.0\linewidth]{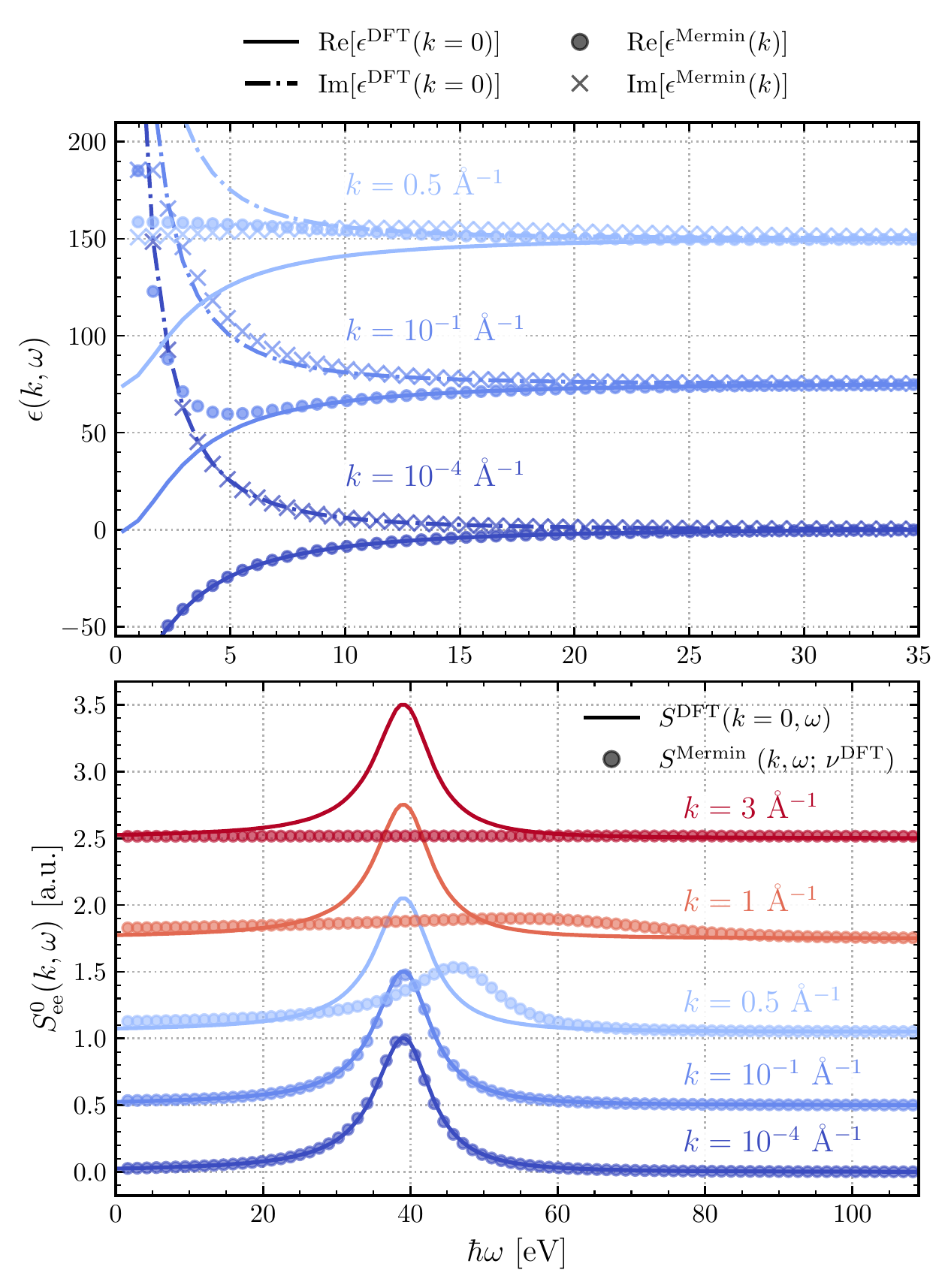}}
\caption{The top panel shows the free-electron part of the dielectric function $\epsilon(k, \omega)$ in a beryllium plasma at $\rho =$~5~g/cm$^3$ and $T=$~100~eV. The DFT results are given at $k=0$, where the solid lines are the real part and the dash-dotted lines are the imaginary part. The Mermin dielectric function from Eq.~\eqref{eq_mermin} is calculated with the DFT collision frequency $\nu^{\mathrm{DFT}}$. The colors represent different values for $k$, while the real and imaginary parts are given by the circles and crosses, respectively. The bottom panel shows the free-electron DSF $S_\mathrm{ee}^0(k,\omega)$ computed from DFT (solid lines) at $k=0$ and from the Mermin dielectric function (circles) at various $k$. The DSFs are scaled to the same magnitude and the dielectric function and DSFs are shifted by 75 and 0.5 a.u., respectively, with respect to the next lowest wave number for readability.}
\label{fig:epslimit}
\end{figure}
\begin{equation} 
  \label{eq_sigmatoeps}
\operatorname{Im} \left[ \epsilon (k=0, \omega ) \right] = \frac{1}{\epsilon_0 \omega} \operatorname{Re} \left[ \sigma(k=0, \omega) \right]
\end{equation}
and use the Kramers-Kronig transformation to compute the corresponding real part, leading to a complex dielectric function $\epsilon^{\mathrm{DFT}}(k=0, \omega)$.
If we require an equivalence between the DFT result and the Mermin dielectric function in the optical limit
\begin{equation} 
  \label{eq_limitfit}
\epsilon^{\mathrm{DFT}} \left(k=0, \, \omega \right) \overset{!}{=} \lim_{k \rightarrow 0} \epsilon^{\mathrm{Mermin}} \left(k, \, \omega; \, \nu \left(\omega \right) \right),
\end{equation}
the real and imaginary parts must be equal simultaneously. This can be achieved by adjusting the real and imaginary part of the dynamic collision frequency which feeds into the Mermin dielectric function, leading to a two dimensional optimization problem. The result of this optimization is a collision frequency $\nu^{\mathrm{DFT}}$ for which the analytic Mermin dielectric function yields the same results as DFT in the macroscopic limit.
Because there is no notion of bound states in the theoretical framework of the Mermin dielectric function, the electrical conductivity must only originate from free or quasi-free states. For this purpose, the conductivity in Eq.~\eqref{eq_KG} can be split into different contibutions, see Ref.~\onlinecite{Bethkenhagen2020} for details.

Figure~\ref{fig:epslimit} shows the convergence of the Mermin dielectric function and DSF to the DFT result in the optical limit for a beryllium plasma at $\rho =$~5~g/cm$^3$ and $T=$~100~eV. Due to the presence of bound states in beryllium at these conditions, only the electrical conductivity due to free electrons can be used as an input to the workflow depicted in Fig.~\ref{fig:flowchart} and all quantities in Fig.~\ref{fig:epslimit} are free-electron contibutions.
The DFT result for the DSF $S^{\mathrm{DFT}}$ and the dielectric function $\epsilon^{\mathrm{DFT}}$ are only available at $k=0$ and are shown as a constant reference for the various $k$ depicted in Fig.~\ref{fig:epslimit}.
In both panels, it is apparent that, with the correct collision frequency $\nu^{\mathrm{DFT}}$, the Mermin result converges to the optical limit described by DFT.
In practice, the limit $k \rightarrow 0$ is reached at wave numbers that correspond to length scales that are significantly larger than any characteristic length scales of the studied system. For beryllium at these conditions, the convergence is reached for wave numbers smaller or equal to $10^{-4}$~\AA$^{-1}$ as depicted in Fig.~\ref{fig:epslimit}. The dielectric functions in the upper panel are connected to the DSF in the lower panel by Eq.~\eqref{eq_FDT}.
However, it is apparent that the dynamic dielectric function in the upper panel of Fig.~\ref{fig:epslimit} is more sensitive to changes in the wave number than the DSF shown in the bottom panel, which is dominated by the pole in $\epsilon^{-1}(k, \omega)$.

\subsection{Linear response time dependent density functional theory}
\label{sec:lrtddft}

In the framework of LR-TDDFT the density response of the non-interacting Kohn-Sham system can be evaluated at a finite momentum transfer as~\cite{Engel2011,Yan2011}:
\begin{multline} 
  \label{eq:LRTDDFT-chiKS}
\chi_\mathrm{KS} (\vec{k}, \omega) = \frac{1}{\Omega} \sum_{\vec{g}, i, j} \frac{f(\epsilon_{i,\vec{g}}) - f(\epsilon_{j,\vec{g} + \vec{k}})}{\omega + \epsilon_{i,\vec{g}} - \epsilon_{j,\vec{g} + \vec{k}} + i \eta } \times \\ \times  \braket{\psi_{i,\vec{g}}| e^{-i \vec{k} \vec{r}} | \psi_{j, \vec{g} + \vec{k}}} \braket{\psi_{i,\vec{g}}| e^{i \vec{k} \vec{r}} | \psi_{j, \vec{g} + \vec{k}}}.
\end{multline}
The quantities in this equation are defined analogously to the Kubo-Greenwood formula in Eq.~\eqref{eq_KG}.
This response function can be related to the full density response $\chi$ via a Dyson equation~\cite{Engel2011}, with different levels of approximation for the exchange-correlation kernel $f_\mathrm{XC}$.
A closed expression can be written as
\begin{equation}
  \label{eq:LRTDDFT-chi}
 \chi (\vec{k}, \omega) = \frac{\chi_\mathrm{KS} (\vec{k}, \omega)}{1 - \left[ v(\vec{k}) + f_\mathrm{XC} (\vec{k}, \omega) \right] \chi_\mathrm{KS} (\vec{k}, \omega)},
\end{equation}
where $v(\vec{k})$ is the Fourier transform of the Coulomb potential. The level of the RPA is achieved for $f_\mathrm{XC} = 0$, for which the dielectric function can be computed as
\begin{equation}
  \label{eq:LRTDDFT-epsRPA}
 \epsilon^\mathrm{RPA}_\mathrm{KS} (\vec{k}, \omega) = 1 - \frac{4 \pi}{\vert \vec{k} \vert^2} \chi_\mathrm{KS} (\vec{k}, \omega).
\end{equation}
Because the Mermin dielectric function accounts for electron interactions on the level of the RPA, we set $f_\mathrm{XC} = 0$ and use Eq.~\eqref{eq:LRTDDFT-epsRPA} in Secs.~\ref{sec:electronDSF-H}, \ref{sec:electronDSF-isoBe} and \ref{sec:electronDSF-compBe} to facilitate comparisons. In Sec.~\ref{sec:experiments}, we use the adiabatic local density approximation~\cite{Zangwill1980,Engel2011}.

\subsection{Computational details}
\label{sec:DFTMD}
All DFT-MD simulations for this work were performed with the Vienna \textit{ab initio} simulation package (VASP)~\cite{Kresse1993, Kresse1994, Kresse1996}. The electronic and ionic parts are decoupled by the Born-Oppenheimer approximation and, for fixed ion positions, the electronic problem is solved in the finite temperature DFT approach~\cite{Mermin1965}. In VASP, the electronic wave functions are expanded in a plane wave basis set up to a energy cutoff $E_\mathrm{cut}$. After the electronic ground state density is determined self-consistently at every time step, the forces on the ions via Coulomb interactions with other ions and the electron cloud are computed and the ions are moved according to Newton's second law. The temperature control in the MD simulation is performed via the Nos\'e{}-Hoover algorithm~\cite{Nose1984, Hoover1985} with a mass parameter corresponding to a temperature oscillation period of 40 time steps. All simulations are performed using the exchange-correlation functional of Perdew, Burke, and Ernzerhof (PBE)~\cite{Perdew1996}.
For beryllium, we use the \verb|PAW_PBE Be_sv_GW 31Mar2010| potential with an energy cutoff of 800~eV for all simulations apart from the compressed case in Sec.~\ref{sec:electronDSF-compBe} for which we use a Coulomb potential with a cutoff of 10~000~eV. For further details on the hydrogen simulation parameters, see Ref.~\onlinecite{Roepke2021}.

The dynamic electrical conductivity, that is the input for the scheme presented in Fig.~\ref{fig:flowchart}, was computed from the eigenfunctions and eigenenergies of separate DFT cycles with a more precise energy convergence criterion via the Kubo-Greenwood formula~\eqref{eq_KG}. These simulations were performed on at least five snapshots taken at equidistant time steps from the DFT-MD simulation. The scheme described in Sec.~\ref{sec:diel} was implemented using the NumPy software package~\cite{numpy2020} for arrays to store the dynamic properties and for the evaluation of simple numerical integration. More elaborate integrals, such as in Eqs.~\eqref{eq_eps_rpa_re3} and \eqref{eq_eps_rpa_im3}, were evaluated using Gaussian quadrature from the SciPy software package~\cite{scipy2020}. The Kramers-Kronig transformation between the real and imaginary part of the dynamic dielectric function and the electrical conductivity was performed according to Maclaurin's formula from Ref.~\onlinecite{Ohta1988}.

The linear response time dependent DFT (LR-TDDFT) calculations were performed in the GPAW code~\cite{Mortensen2005,Enkovaara2010,Larsen2017,Yan2011}. The same snapshots as for the Kubo-Greenwood calculations were used and a 2x2x2 or 4x4x4 Monkhorst-Pack grid~\cite{Monkhorst1976} was employed for calculations of $k$-dependent dielectric functions. For the considered conditions, already the Baldereschi mean value point~\cite{Baldereschi1973} yields converged optical conductivities for the Kubo-Greenwood calculations. For hydrogen, the dielectric function was computed with a plane-wave energy cutoff of at least 50~eV, while for beryllium at least 250~eV were used.

\section{Dynamic collision frequency}
\label{sec:dynColl}

\begin{figure}[t]
\center{\includegraphics[angle=0,width=1.0\linewidth]{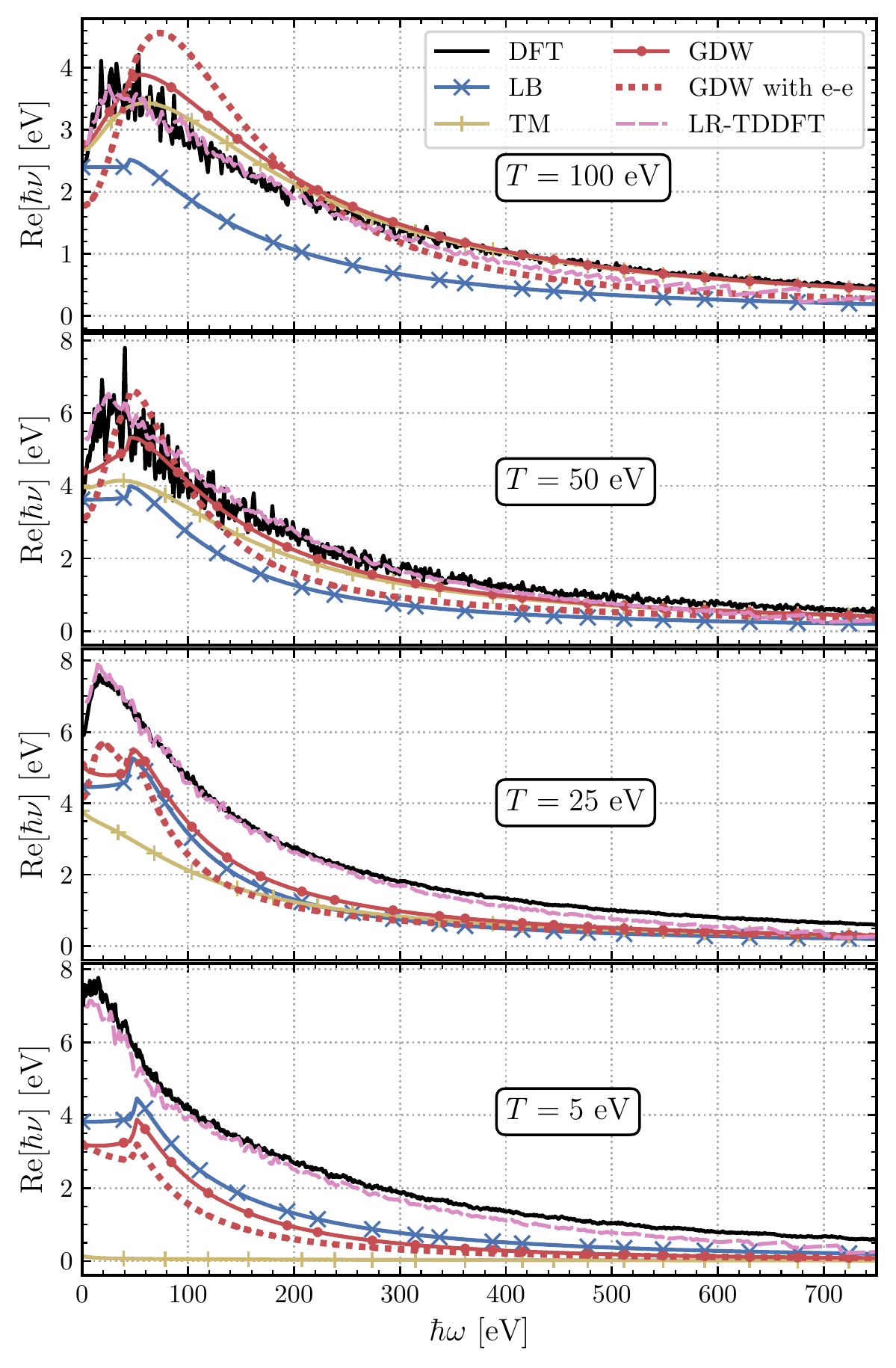}}
\caption{The real part of the dynamic collision frequency of hydrogen plasmas at $\rho = 2$~g/cm$^3$ for temperatures ranging from $5$ to $100$~eV. The DFT and LR-TDDFT collision frequencies determined via Eq.~\eqref{eq_limitfit} from their respective electrical conductivities are shown in black and pink, respectively. The LB collision frequency is shown in blue with crosses and the $T$-Matrix approach is shown in yellow with plus symbols. The GDW collision frequencies with and without electron-electron collisions are depicted in red as a dotted line and as a solid line with filled circles, respectively.}
\label{fig:CollCompFull}
\end{figure}
The work flow presented in Fig.~\ref{fig:flowchart} results in a complex dynamic collision frequency $\nu^\mathrm{DFT} (\omega)$. To study how this collision frequency compares to different levels of analytic approximations, we determine the real part of $\nu^\mathrm{DFT}$ for a hydrogen isochore at $\rho=2$~g/cm$^3$ from 5 to 100~eV (see Ref.~\onlinecite{Roepke2021} for numerical details). This temperature range was chosen to illustrate the transition from the WDM regime to the ideal plasma regime.
In Fig.~\ref{fig:CollCompFull} we compare these collision frequencies to the Lenard-Balescu (LB) collision frequency, the $T$-Matrix (TM) approach and the Gould-DeWitt (GDW) approach. The LB approach goes beyond the Born collision frequency by including dynamic screening, while the TM approach accounts for strong binary collisions by summing up ladder diagrams in the perturbation expansion~\cite{Reinholz2005}. The GDW scheme combines the dynamic screening of the LB approach with the strong collisions of the TM treatment and should, in principle, give the most accurate results. For further details on the analytic approaches see Refs.~\onlinecite{Reinholz2005,Thiele2006,Reinholz2000,Gould1967,Roepke1989}. The aforementioned approaches solely describe electron-ion collisions, but electron-electron (e-e) collisions can be included by modulating the collision frequency with a renormalization factor~\cite{Reinholz2000}. The GDW collision frequency including e-e collisions is also indicated in Fig.~\ref{fig:CollCompFull} by the red dotted lines. It is apparent that although the DFT predictions agree well with the TM and GDW approach at high temperatures, it deviates significantly at lower temperatures where complex many-body and quantum effects contribute strongly. At $T=100$~eV, the collision frequency is dominated by strong collisions between ions and electrons. However, the inclusion of e-e collisions via the renormalization factor leads to worse agreement with the DFT results, which is in agreement with recent observations that the Kubo-Greenwood formula applied to DFT lacks e-e collisions~\cite{Roepke2021,French2022}. Furthermore, we apply the work flow presented in Fig.~\ref{fig:flowchart} to the electrical conductivity in the optical limit computed by LR-TDDFT to extract a collision frequency which we show as the pink dashed lines in Fig.~\ref{fig:CollCompFull}.
At all temperatures, its behavior is very similar to the Kubo-Greenwood results which indicates that electron-electron collisions are also not included in this description of transport properties. It is remarkable that at high frequencies the LR-TDDFT collision frequency lies significantly below the Kubo-Greenwood results for all considered temperatures. In our tests, this could not be attributed to a lack of convergence in number of bands or cutoff energy.

\section{Dynamic structure factor}
\label{sec:electronDSF}

\subsection{Hydrogen}
\label{sec:electronDSF-H}

\begin{figure}[b] 
\center{\includegraphics[angle=0,width=1.0\linewidth]{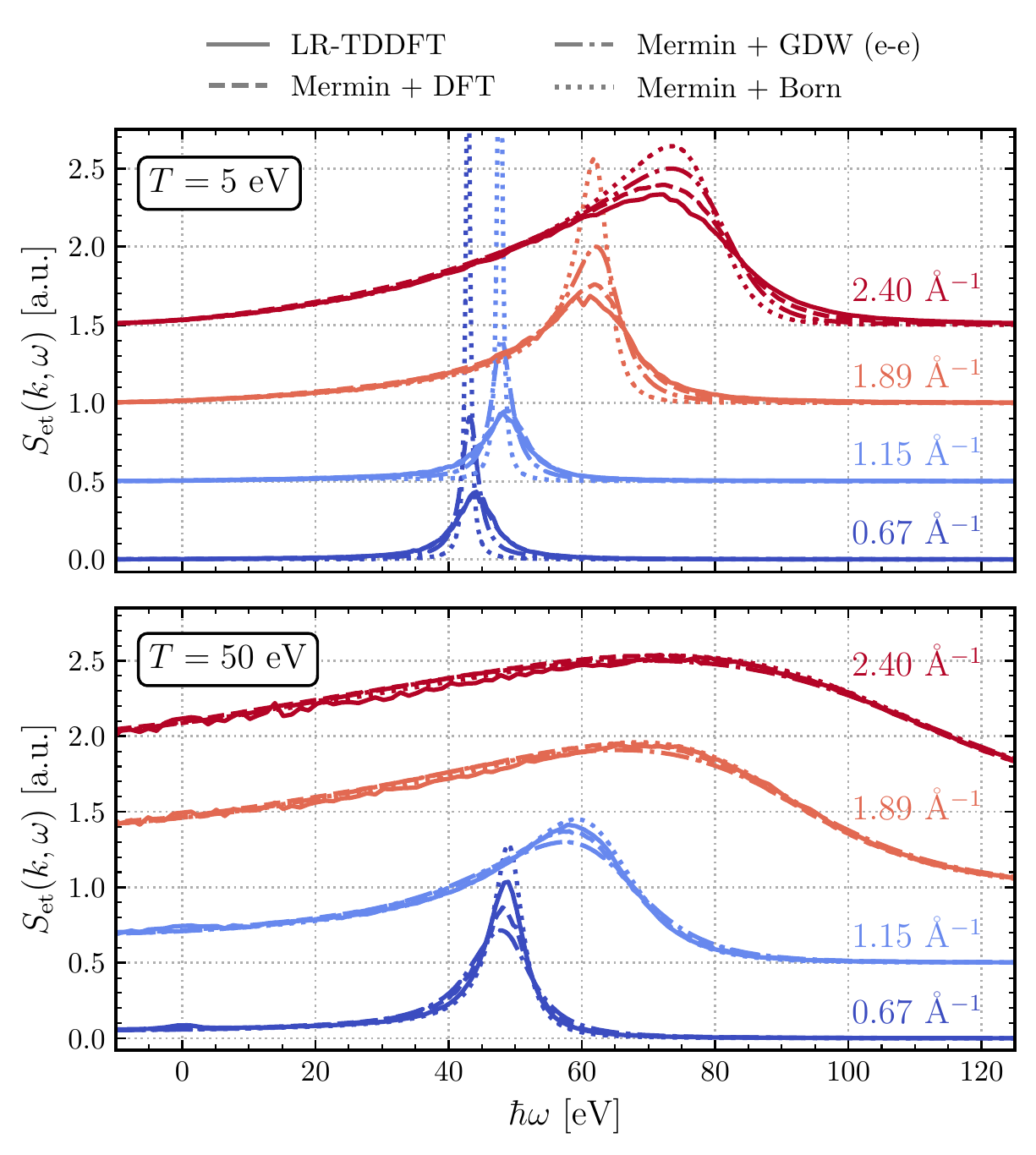}}
\caption{The inelastic electronic DSF $S_{\mathrm{et}} (k, \, \omega)$ of a hydrogen plasma at $\rho = 2$~g/cm$^3$ and $T=5$~eV (upper panel) and $T=50$~eV (lower panel) from $k=0.67$~\AA$^{-1}$ to $k=2.40$~\AA$^{-1}$. The solid line denotes the direct computation from LR-TDDFT at the respective wave numbers, while the other lines denote DSFs computed from the Mermin dielectric function with the DFT collision frequency (dashed lines), the GDW collision frequency including electron-electron collisions (dash-dotted lines) and the Born collision frequency (dotted lines). The DSFs are shifted by 0.5 with respect to the next lowest wave number for readability.}
\label{fig:H50eVDSF}
\end{figure}
Given a dynamic collision frequency $\nu (\omega)$, Eqs.~\eqref{eq_FDT} and \eqref{eq_mermin} can be used to compute the electronic DSF $S_{ee} (k, \, \omega)$ where the $k$ dependence only enters through the Mermin dielectric function. The LR-TDDFT approach allows direct access to the dielectric function at finite $k$ by computing transitions matrix elements between Kohn-Sham states at different $k$ points~\cite{Yan2011}.
In Fig.~\ref{fig:H50eVDSF}, we show the electronic DSF of a hydrogen plasma at $\rho = 2$~g/cm$^3$ and $T=50$~eV (lower panel) and $T=5$~eV (upper panel). The direct computations through LR-TDDFT are shown as solid lines, while we also present DSFs computed via the Mermin dielectric function in conjunction with the DFT and GDW collision frequencies shown in Fig.~\ref{fig:CollCompFull} as dashed and dash-dotted lines, respectively. Additionally, we show the results from the Mermin dielectric function with the Born collision frequency (see Eq.~\eqref{eq_born_coll}), which constitutes the often used Born-Mermin approach, as dotted lines. At the lowest wave number shown in Fig.~\ref{fig:H50eVDSF}, $k=0.67$~\AA$^{-1}$, we are considering the collective behavior where collision are important, as can be seen from the dimensionless scattering parameter $\alpha$ (see Ref.~\onlinecite{Glenzer2009} for definition) which is 4.17 and 2.84 for $T=5$ and $T=50$~eV, respectively.

As expected for a fully ionized hydrogen plasma, the $k$ dependence encoded by the Mermin dielectric function agrees well with the direct computation via LR-TDDFT for all considered collision frequencies at both conditions. However, at $T=5$~eV, the damping of the plasmon predicted by LR-TDDFT can only be captured with the DFT collision frequency, especially at small $k$. The Born collision frequency leads to a vast overestimation of the plasmon magnitude for $k$ below $2.4$~\AA$^{-1}$ and also the GDW approach with renormalization overestimates the magnitude by a factor of 2 for $k$ below $1.15$~\AA$^{-1}$. With increasing wave numbers, the collisions become less significant, and the DSFs for all collision frequencies start to converge to the same result.
At $T=50$~eV, the collisions play a smaller role, which is demonstrated by the largely identical predictions from all collision frequencies for $k$ above $1.15$~\AA$^{-1}$. It is notable that although the inclusion of electron-electron collisions leads to significant discrepancies between the dynamic collision frequencies in Fig.~\ref{fig:CollCompFull}, these differences cannot be observed in the DSF, given the numerical noise.
In the LR-TDDFT data, a small additional contribution at $\hbar \omega = 0$~eV appears, which has also recently been seen in path integral Monte Carlo simulations~\cite{Dornheim2022}. This bump is not included in the Mermin formalism and appears more pronounced at higher temperatures and lower densities (also see Sec.~\ref{sec:electronDSF-isoBe} and \ref{sec:electronDSF-compBe}), leading us to propose that it is connected to bound-bound transitions without energy transfer.

\subsection{Isochorically heated beryllium}
\label{sec:electronDSF-isoBe}

\begin{figure}[b]
\center{\includegraphics[angle=0,width=1.0\linewidth]{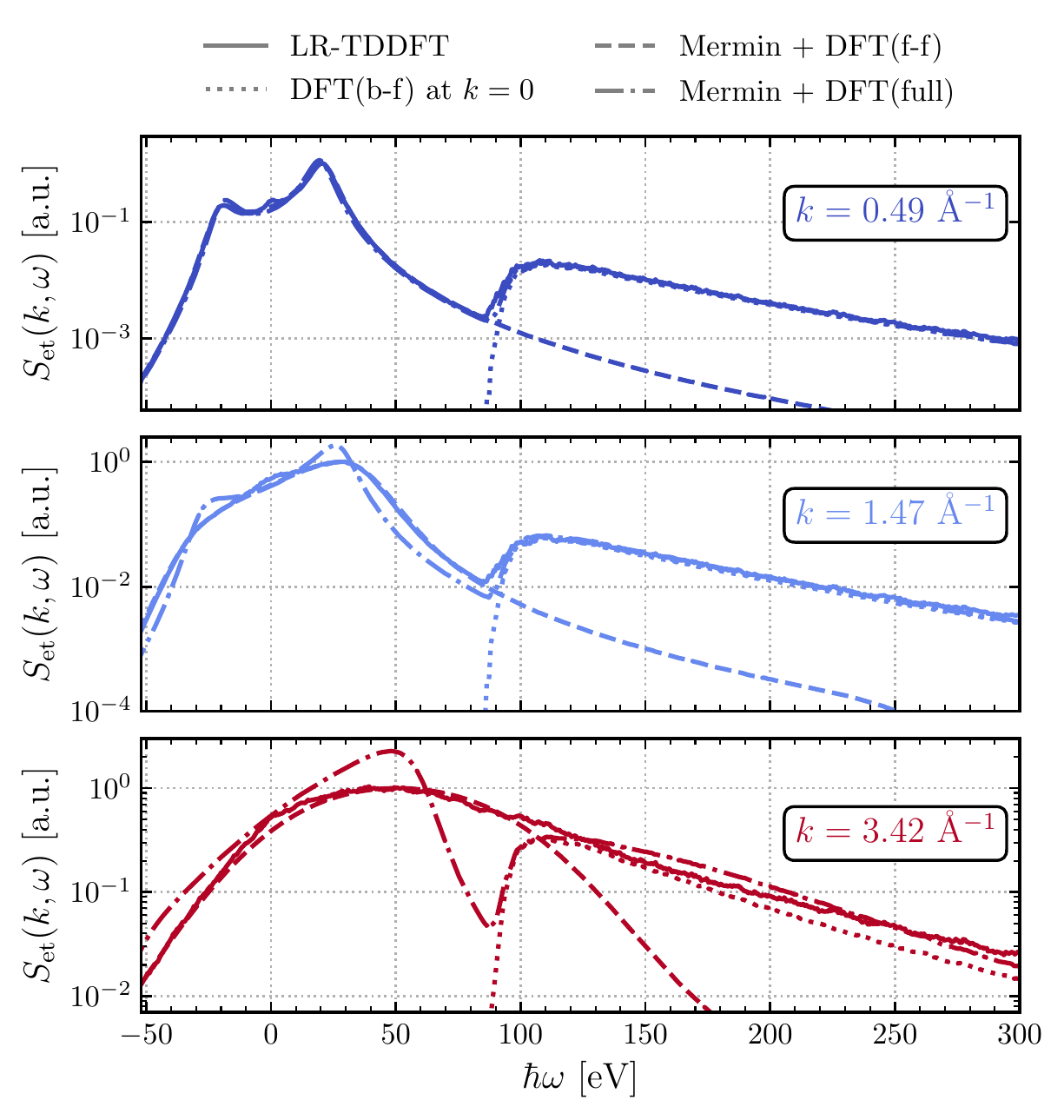}}
\caption{The inelastic electronic DSF $S_{\mathrm{et}} (k, \, \omega)$ of a beryllium plasma at $\rho = 1.8$ ~g/cm$^3$ and $T=12$~eV for various $k$ values on a logarithmic scale. The solid lines are direct computations at the given $k$ using LR-TDDFT. The dash-dotted and the dashed lines denote DSFs computed from the Mermin dielectric function with the full DFT collision frequency, determined from the electrical conductivity including bound-free transitions, and the free-free collision frequency, determined from the electrical conductivity including only free-free transitions, respectively. The dotted lines denote the DSF computed directly from the bound-free conductivity at $k=0$~\AA$^{-1}$.}
\label{fig:Be12eVDecomp}
\end{figure}

To investigate the impact of tightly bound states on the presented procedure, we study a beryllium plasma at $\rho = 1.8$~g/cm$^3$ and $T=12$~eV, for which the approach of Ref.~\onlinecite{Bethkenhagen2020} predicts a charge state $Z=2.1$. The bound 1s states are energetically clearly separated from the free electrons. The collision frequency can either be determined from the full dynamic electrical conductivity that includes the transitions from the bound 1s states to the conduction band, or from the free-free electrical conductivity by restricting the transition matrix elements in Eq.~\eqref{eq_KG} to transitions orginating and ending in the conduction band (for details on this decomposition, see Ref.~\onlinecite{Bethkenhagen2020}). In the latter case, only the free-free contribution to the DSF is considered within the Mermin dielectric function, while the bound-free contribution must be approximated by its behavior at $k \rightarrow 0$.
In Fig.~\ref{fig:Be12eVDecomp}, we show the comparison of these two approaches to the direct computation of the electronic DSF using LR-TDDFT. At the lowest wave number $k = 0.49$~\AA$^{-1}$, shown in the upper panel, all approaches agree well, as expected due to the construction of the collision frequency which requires equivalence in the limit of small $k$ (see Eq.~\eqref{eq_limitfit}). The separation of the conductivity into a free-free and a bound-free contribution allows us to clearly identify the different terms of the Chihara formula~\eqref{eq_chihara} in the DSF. The dotted line represents the bound-free contribution, which agress exactly with the LR-TDDFT data above $\sim 90$~eV, and the dashed line represents the free-free contribution (plasmon), which matches the LR-TDDFT results below $\sim 90$~eV. Remarkably, the prefactors $Z_\mathrm{f}$ and $Z_\mathrm{b}$ in Eq.~\eqref{eq_chihara} which give the respective weighting of these two features come out of the definition of the charge state described in Ref.~\onlinecite{Bethkenhagen2020} and agree virtually exactly with the direct computation including all transitions in LR-TDDFT.

\begin{figure}[t]
\center{\includegraphics[angle=0,width=1.0\linewidth]{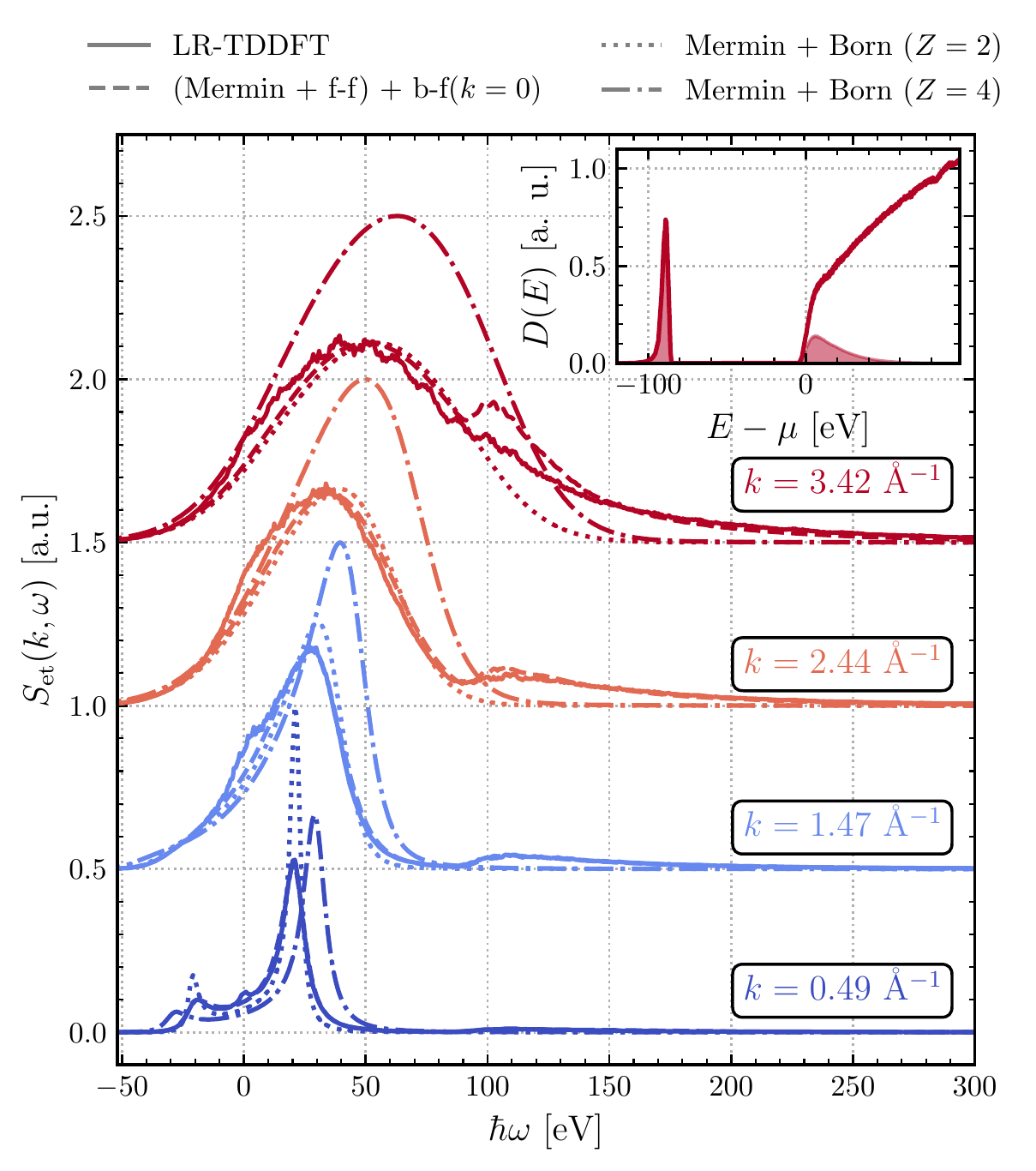}}
\caption{The inelastic electronic DSF $S_{\mathrm{et}} (k, \, \omega)$ of a beryllium plasma at $\rho = 1.8$ ~g/cm$^3$ and $T=12$~eV for various $k$ values. The solid lines are direct computations at the given $k$ using LR-TDDFT, while the dotted and dash-dotted lines denote DSFs computed from the Mermin dielectric function with the Born collision frequency for a plasma with a charge state of $Z=2$ and $Z=4$, respectively. The dashed lines represent the sum of the DSF computed through the Mermin dielectric function using the free-free collision frequency and the bound-free DSF at $k=0$~\AA$^{-1}$. The DSFs are shifted by 0.5 with respect to the next lowest wave number for readability. In the inset, the solid line shows the density of states, while the shaded area denotes the occupied density of states.}
\label{fig:Be12eVDSF}
\end{figure}

At $k = 1.47$~\AA$^{-1}$ in the middle panel of Fig.~\ref{fig:Be12eVDecomp}, the deviation of the approach using the full collision frequency to the other approaches becomes apparent. The bound-free dominated DSF above $\sim 90$~eV is still well approximated by both the full collision frequency and the bound-free feature at $k \rightarrow 0$. Below $\sim 90$~eV, however, the approach using the full collision frequency, denoted by the dash-dotted line, deviates strongly (note the logarithmic scale) from the LR-TDDFT result. The free-free feature computed solely from the collision frequency based on free-free transitions, denoted by the dashed line, still agrees very well with the LR-TDDFT calculation in this energy regime. 

The bottom panel of Fig.~\ref{fig:Be12eVDecomp}, showing the DSF at $k = 3.42$~\AA$^{-1}$, highlights the complete breakdown of the approach using the full collision frequency. While the DSF is still described adequately above $\sim 90$~eV, its shape is very different from the LR-TDDFT result below that energy. On the other hand, the separate description of free-free and bound-free contributions again describes the DSF accurately compared to the LR-TDDFT data. However, the approximation of the bound-free feature by its $k \rightarrow 0$ limit starts to deteriorate at this wave number. At the highest energy shift shown in Fig.~\ref{fig:Be12eVDecomp}, this approximation underestimates the LR-TDDFT value by a factor of almost 2. Additionally, at the onset of the bound-free feature around $100$~eV, it overestimates the DSF compared to the LR-TDDFT as can be seen in Fig.~\ref{fig:Be12eVDSF} which shows the DSF on a linear scale. The fast deterioration beyond the $k \rightarrow 0$ limit of the approach using the full collision frequency is expected because the framework of the Mermin dielectric function, which encodes the $k$ dependence, does not include the existence of bound states. Therefore, any such states that are artificially introduced via the collision frequency cannot be handled correctly in the $k$ dependence.

Furthermore, in Fig.~\ref{fig:Be12eVDSF}, we show the DSFs computed from the Mermin dielectric function with Born collision frequencies for a plasma with a charge state $Z=2$ and $Z=4$. The position of the plasmon peak for $Z=2$ agrees well with the DFT spectra, while the position of the $Z=4$ plasma is consistently at too high energies, as expected due to the higher free-electron density. However, at low $k$, the dampening of the plasmon peak due to the Born collision frequency is too low compared to the DFT data, similar as observed for hydrogen in Fig.~\ref{fig:H50eVDSF}. At the higher wave numbers, the plasmon-peak position of the DFT results agrees well with Mermin function using the Born collision frequency at $Z=2$, clearly indicating that the bound 1s states do not contribute to this feature. The inset in Fig.~\ref{fig:Be12eVDSF} shows the density of states (DOS) of the beryllium plasma which shows a clear separation between the narrow 1s band, which is fully occupied, and the conduction band. This clear distinction is the reason why the separate treatment of free-free and bound-free contibutions is successful. The bound-free feature does not exhibit a strong $k$ dependence up to high $k$ values~\cite{Mattern2013,Eisenberger1970}, and the plasmon occurs energetically separated in the DSF.

\subsection{Compressed beryllium}
\label{sec:electronDSF-compBe}

\begin{figure}[t]
\center{\includegraphics[angle=0,width=1.0\linewidth]{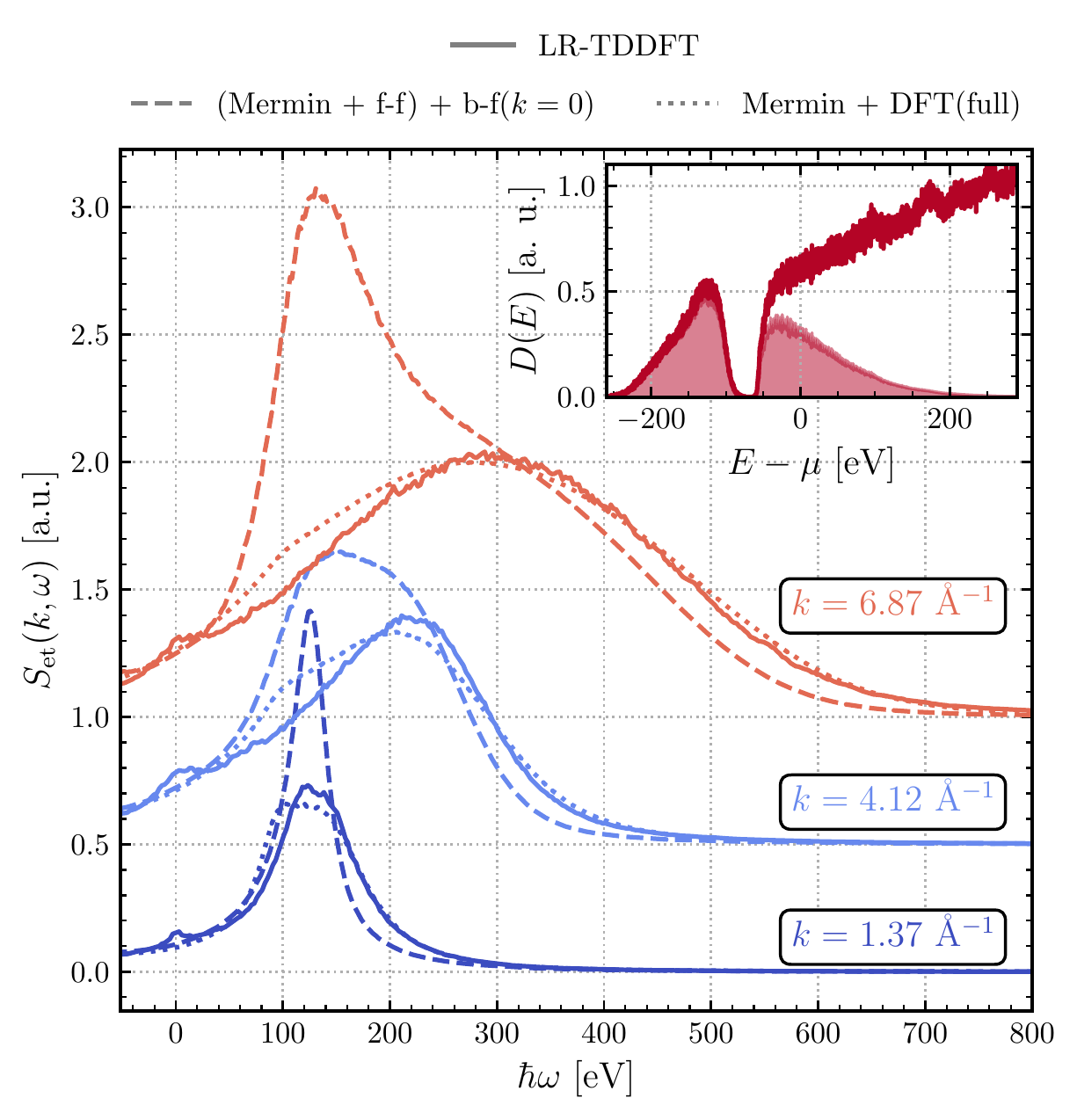}}
\caption{The inelastic electronic DSF $S_{\mathrm{et}} (k, \, \omega)$ of a beryllium plasma at $\rho = 40$ ~g/cm$^3$ and $T=50$~eV for various $k$ values. The solid lines are direct computations at the given $k$ using LR-TDDFT, while the dashed lines represent the sum of the DSF computed through the Mermin dielectric function using the free-free collision frequency and the bound-free DSF at $k=0$~\AA$^{-1}$. The dotted lines denote the DSF computed through the Mermin dielectric function with the full collision frequency. The DSFs are shifted by 0.5 with respect to the next lowest wave number for readability. In the inset, the solid line shows the density of states, while the shaded area denotes the occupied density of states.}
\label{fig:Be50eVDSF}
\end{figure}

With increasing density and temperature the notion of bound states becomes ill-defined in WDM. The inset in Fig.~\ref{fig:Be50eVDSF} shows the DOS of a beryllium plasma at $T=50$~eV and $\rho=40$~g/cm$^3$ which demonstrates the closing of the band gap compared to the inset in Fig.~\ref{fig:Be12eVDSF}. Furthermore, the former 1s states broaden significantly into a band and the DOS converges towards the $\sqrt{E}$ behavior of a free electron gas.
Because the band gap is still clearly identifiable the separate treatment of bound-free and free-free contributions to the DSF presented in the previous section can also be applied to these conditions. Figure~\ref{fig:Be50eVDSF} shows the results of this separate treatment, as well as the direct computation using LR-TDDFT and the DSF from the Mermin dielectric function using the full collision frequency.
While the separate treatment of bound-free and free-free contributions yields excellent results for the near-ambient density case in Fig.~\ref{fig:Be12eVDecomp}, it poorly approximates the LR-TDDFT results in strongly compressed beryllium shown in Fig.~\ref{fig:Be50eVDSF}. The plasmon peak at $k=1.37$~\AA$^{-1}$ is severly underdamped due to the missing bound-free transitions in the collision frequency, which occur in the same energy range as the free-free transitions at these conditions. The use of the Born collision frequency in lieu of the free-free DFT collision frequency leads to an increase of the plasmon peak magnitude by a factor of 2 (not shown in Fig.~\ref{fig:Be50eVDSF}). The broader peak arising around $\sim 130$~eV for $k=4.12$ and $k=6.87$~\AA$^{-1}$ is due to the insufficient approximation of the bound-free feature by its value at $k=0$~\AA$^{-1}$. As can be seen from the LR-TDDFT data, the bound-free features merges with the free-free feature to form one homogeneous feature.
At these conditions, using the full collision frequency in the Mermin dielectric function gives better results, which is expected as the former 1s states lose their bound character due to the higher compression and higher temperature. For all considered wave numbers, this approach yields good agreement with the LR-TDDFT data above $\sim 200$~eV, and approximates the trends below that energy fairly well. Solely at $\sim 100$~eV this approach predicts a feature that is not visible in the LR-TDDFT results across the considered $k$ range.

\section{Application to experiments}
\label{sec:experiments}

\begin{figure}[b]
\center{\includegraphics[angle=0,width=1.0\linewidth]{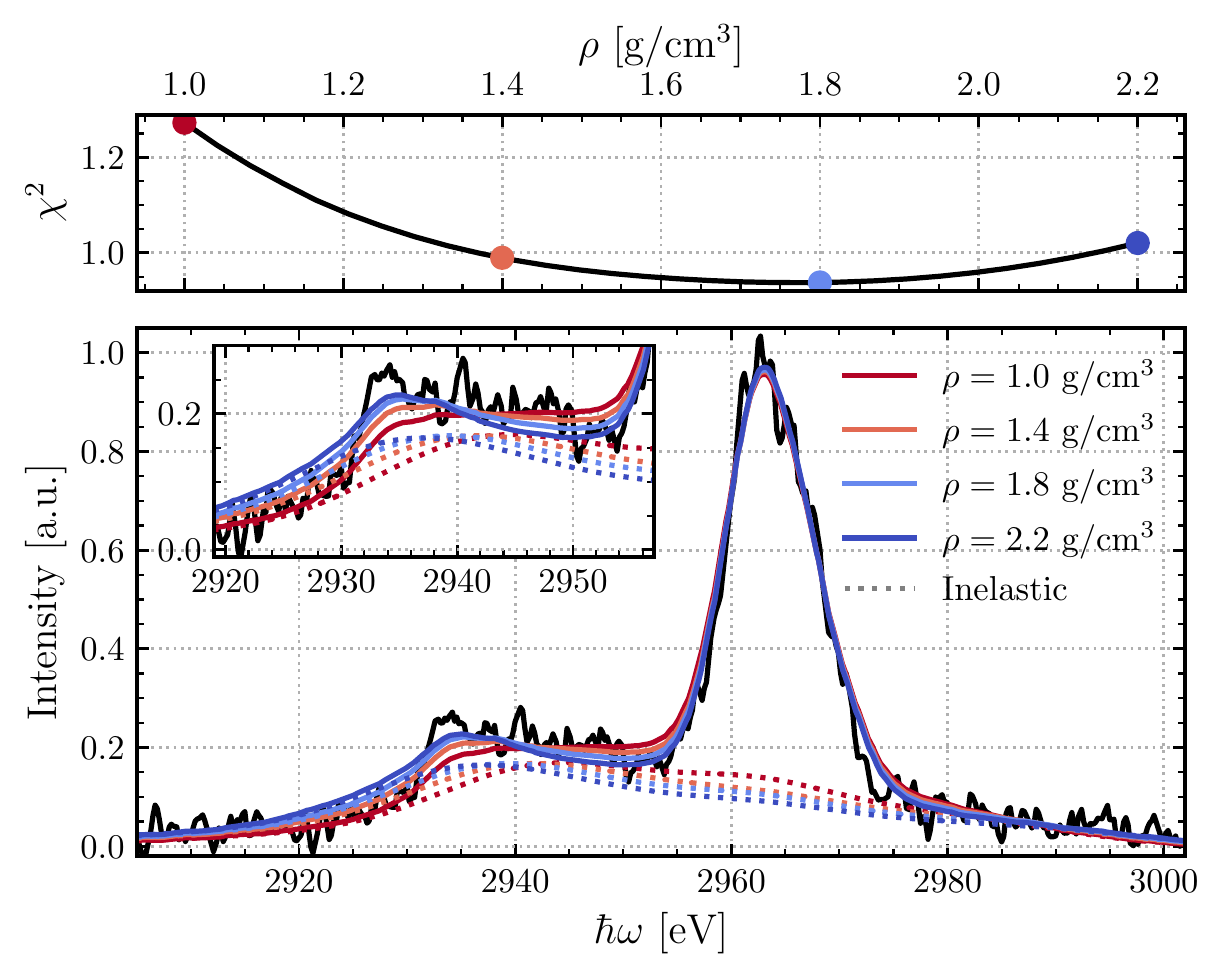}}
\caption{The lower panel shows the scattering intensity of an isochorically heated beryllium target at $T=18$~eV from Ref.~\onlinecite{Doeppner2009}. The colors of the solid lines encode different densities used in the LR-TDDFT simulations. The dotted lines denote the inelastic contributions. The upper panel shows the $\chi^2$ deviation depending on the density used in the simulation where the colored dots correspond to the spectra shown in the lower panel and the black curve is achieved by interpolating to 40 evenly spaced densities between these spectra.}
\label{fig:DoeppnerHEDP}
\end{figure}
We reanalyze previous XRTS experiments by Döppner~\textit{et~al.}~\cite{Doeppner2009} and Kritcher~\textit{et~al.}~\cite{Kritcher2011} using LR-TDDFT to evaluate the influence of advanced methods on the initially inferred plasma parameters. In general, temperature and density of the target must be considered simultaneously. However, since Döppner~\textit{et~al.} used detailed balance in their forward scattering experiment to determine the temperature as $T=18$~eV, we use this value and vary the density to find the best agreement with the experimental data. To justify this approach we show the results of a recently suggested model-free temperature diagnostic~\cite{Dornheim2022_natcom} in Appendix~\ref{sec:app-laplace}. For the other experiment, we include the temperature in the analysis.

Firstly, in Fig.~\ref{fig:DoeppnerHEDP}, we show simulated XRTS spectra with densities ranging from $1.0$ to $2.2$~g/cm$^3$ at $T=18$~eV together with the forward XRTS spectrum recorded by Döppner~\textit{et~al.}~\cite{Doeppner2009}, which was collected at the Omega laser facility at the Laboratory for Laser Energetics at the University of Rochester. The experiment probed a scattering vector of approximately $k=1$~\AA$^{-1}$, enabling access to collective behavior of the plasma. In the original analysis of the experiment a density of $1.17$~g/cm$^3$ was determined by Döppner~\textit{et~al.}~\cite{Doeppner2009}. The electron feature was treated on the level of the RPA wihtout including electron-ion collisions and the ionization was assumed to be $Z_\mathrm{f}=2.3$.
We compute the electron feature for various densities from LR-TDDFT while including local field corrections via the adiabatic local density approximation~\cite{Zangwill1980,Engel2011}. The magnitude of the ion feature is left as a free parameter in the $\chi^2$ minimization. Although none of the computed spectra capture the plasmon at $2930$~eV perfectly, the spectrum at $\rho = 1.8$~g/cm$^3$ yields a 5\% lower $\chi^2$ deviation than any of the other considered densities.
The ionization state at this density is $Z=2.14$, determined via the Thomas-Reiche-Kuhn sum rule~\cite{Bethkenhagen2020}, which is approximately 7\% lower than the value used by Döppner~\textit{et~al.}~\cite{Doeppner2009}. Furthermore, the here computed density of the sample is more than 50\% higher than the originally determined value, indicating the need to use sophisticated methods to achieve reliable results in forward scattering experiments.

\begin{figure}[t]
\center{\includegraphics[angle=0,width=1.0\linewidth]{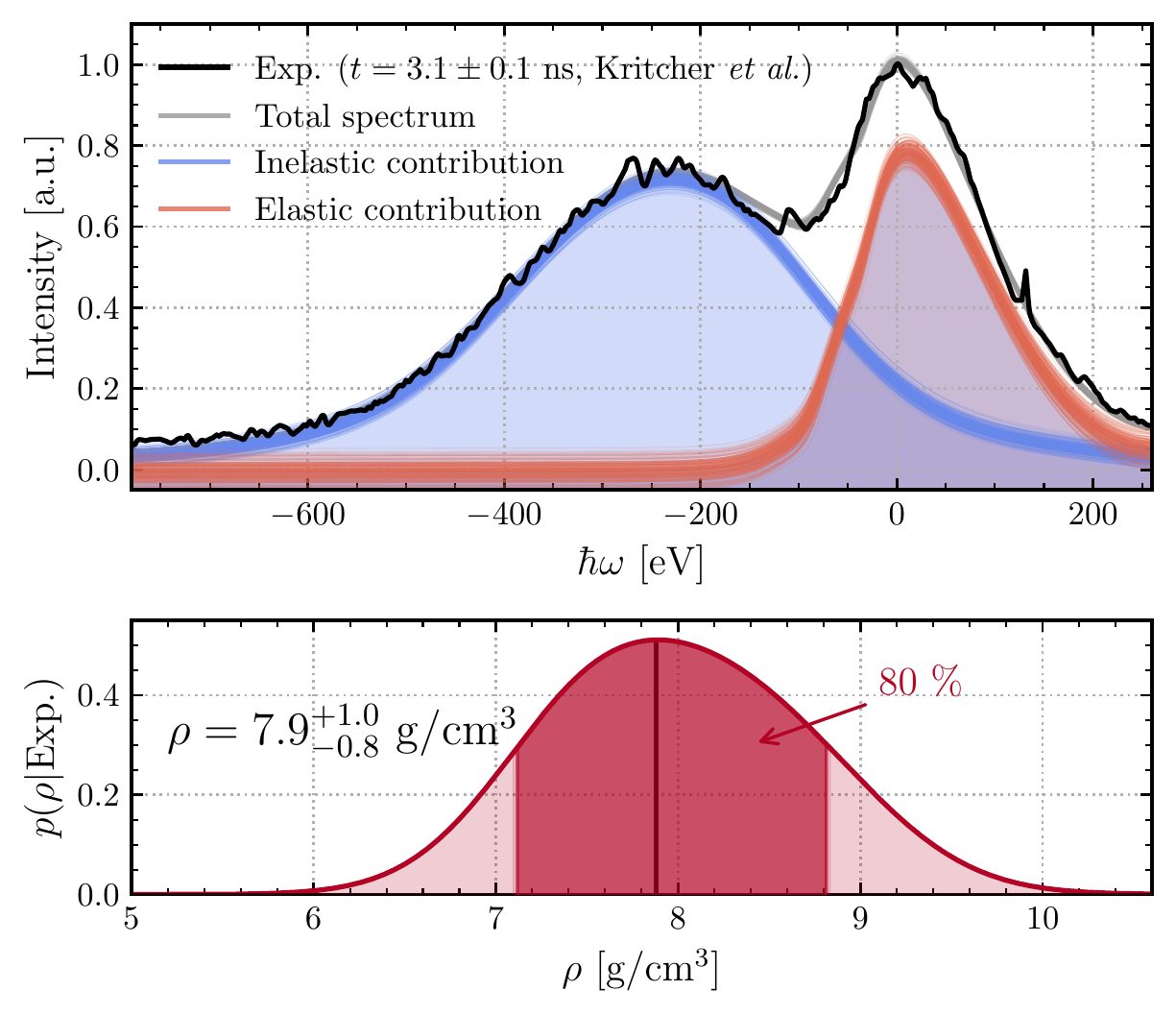}}
\caption{Scattering intensity of imploding beryllium shells from Ref.~\onlinecite{Kritcher2011}. The upper panel shows the experimental data at a delay $t=3.1 \pm 0.1$~ns and the posterior prediction for the elastic and inelastic contributions based on LR-TDDFT simulations. The thin lines are 100 spectra computed from parameters randomly sampled from the posterior probility distribution. The shaded areas show the region below the average posterior predictions. The lower panel shows the reduced posterior probability distribution in the density parameter $\rho$ where the dark shaded area under the curve indicates the 80\% highest posterior density interval.}
\label{fig:KritcherPRL}
\end{figure}

For the experiment by Kritcher~\textit{et~al.}~\cite{Kritcher2011}, the temperature cannot reliably be inferred from the detailed balance relation and the temperature must, therefore, be included in the analysis. Furthermore, the instrument and source functions were not available and must be modeled explicitly in the analysis. To analyze the experiments, we simulate spectra on a sufficiently large temperature and density grid and interpolate between them~\cite{Weissker2009} to model arbitrary $\rho-T$ combinations in this range. Due to the high number of parameters involved in this sort of analysis, we employ Bayesian inference~\cite{Kasim2019} implemented in the PyMC3 software package~\cite{Salvatier2016} and use the sequential Monte Carlo algorithm~\cite{DelMoral2006,Sisson2007} for sampling the parameter space. In Fig.~\ref{fig:KritcherPRL}, we consider the backward XRTS experiment at $k = 8.42$~\AA$^{-1}$ on imploding beryllium shells by Kritcher~\textit{et~al.}~\cite{Kritcher2011}, which was also performed at the Omega laser facility.
To analyze the experiment, we simulate spectra on a grid ranging from $2$ to $32$~g/cm$^3$ and from $0.1$ to $25$~eV. No instrument or source function was supplied in Ref.~\onlinecite{Kritcher2011}. We, therefore, use the parametrization of a zinc source given in Ref.~\onlinecite{MacDonald2021} and include all the parameters of the instrument response function in the Bayesian analysis. We also replace the Gaussian describing the source broadening by a skewed Gaussian to account for the asymmetry observed in the ion feature. Thus, ten parameters determine the shape of the spectrum, including the physical parameters describing the density and temperature of the sample and the magnitude of the ion feature, and 7 parameters describing the experimental setup. The upper panel of Fig.~\ref{fig:KritcherPRL} shows an XRTS spectrum collected from an imploding beryllium shell at a delay of $t=3.1 \pm 0.1$~ns and the posterior prediction for the elastic and inelastic contribution to the simulated scattering spectrum. The posterior predictions are obtained by sampling parameters according to the posterior probability distribution and using these parameters to simulate the spectrum.
The agreement between the simulated spectrum and the experimental data is excellent. The bottom panel of Fig.~\ref{fig:KritcherPRL} shows the reduced posterior probability distribution in the density parameter $\rho$, which is the full probability distribution integrated over all other parameters. The inferred density $\rho=7.9^{+ 1.0}_{- 0.8}$~g/cm$^3$ corresponds to the maximum aposteriori probability and the uncertainties are determined from the 80\% highest posterior density interval. With an assumed ionization state $Z=2$, the original analysis by Kritcher~\textit{et~al.}~\cite{Kritcher2011} resulted in estimates of $\rho = 8.23 \pm 2.24$~g/cm$^3$ and $T=14 \pm 3$~eV. The density, which is the most sensitive plasma parameter with respect to the Compton feature at these conditions, agrees very well with our current study. However, Kritcher~\textit{et~al.} also used a temperature-dependent model for the ion feature, while we keep the ion feature as a free parameter. Therefore, the inferred temperature is mainly determined from the relative magnitude of the ion feature and Compton feature. Because the shape of the Compton feature is not very sensitive to the temperature at these conditions, we cannot reliably determine the electron temperature.


\section{Conclusion}
\label{sec:concl}
In this work, we presented the theoretical basis for computation of DSFs using the Mermin dielectric function with a dynamic complex collision frequency and showed how this framework can be used to extract collision frequencies from DFT simulations. We compared these collision frequencies to several analytic approaches for hydrogen plasmas at $\rho=2$~g/cm$^3$ and, for temperatures approaching the ideal plasma limit, found good agreement with models that incorporate strong collisions. Furthermore, we studied how different collision frequencies impact the DSF calculated from the Mermin dielectric function and compared these results to the direct computation of the DSF at the given wave numbers using LR-TDDFT. For hydrogen, we find good agreement for all collision frequencies at high $k$, while at small $k$, especially the frequently used Born approximation leads to underdamped plasmon peaks. For beryllium, we showed that a separate treatment of free-free and bound-free contributions to the DSF yields excellent agreement with the LR-TDDFT for near-ambient densities up to moderate wave numbers ($k=3.42$~\AA$^{-1}$), while it disagrees significantly for highly compressed beryllium because bound-free transition interact with the free-free transitions to dampen the plasmon. Therefore, in order to get accurate DSFs over a wide range of wave numbers in extreme conditions, it is imperative to employ \textit{ab~initio} approaches like LR-TDDFT or path integral Monte Carlo simulations.
We applied LR-TDDFT to XRTS experiments on beryllium and found significant deviations of more than 50\% in inferred density for small $k$ for Döppner~\textit{et~al.}~\cite{Doeppner2009} and found good agreement with analytical approaches for backscattering with large $k$ for Kritcher~\textit{et~al.}~\cite{Kritcher2011}.

\begin{acknowledgments}
We want to thank P. Sperling, B. Witte, M. French, G. Röpke, H.~J. Lee and A. Cangi for many helpful discussions. M.~S. and R.~R. acknowledge support by the Deutsche Forschungsgemeinschaft (DFG) within the Research Unit FOR 2440. All simulations and analyses were performed at the North-German Supercomputing Alliance (HLRN) and the ITMZ of the University of Rostock. M.~B. gratefully acknowledges support by the European Horizon 2020 programme within the Marie Skłodowska-Curie actions (xICE grant 894725) and the NOMIS foundation. The work of T.~D. was performed under the auspices of the U.S. Department of Energy by Lawrence Livermore National Laboratory under Contract No. DE-AC52-07NA27344.
\end{acknowledgments}

\appendix

\section{Derivation of real and imaginary part of RPA dielectric function}
\label{sec:app-rpa}

The collision frequency is generally a complex number
\begin{equation} 
  \label{eq_nu}
\nu(\omega) = \nu_1(\omega) + i \, \nu_2(\omega) ,
\end{equation}
meaning that its imaginary part acts as a shift of the frequency that enters into $\epsilon^\mathrm{RPA}$ in Eq.~\eqref{eq_mermin} and its real part takes on the role of the artificial damping $\eta$ that was introduced in Eq.~\eqref{eq_eps_rpa}. However, in this case, the damping is not set to zero after the integration.

Now, we will split Eq.~\eqref{eq_eps_rpa} into its real and imaginary part and consider the modulation of the input frequency $\omega$ by the complex frequency from Eq.~\eqref{eq_nu} where the argument of $\nu$ is dropped for readability:
\begin{multline} 
  \label{eq_eps_rpa_re}
\operatorname{Re} \left[ \epsilon^\mathrm{RPA} (\vec{k}, \omega + i \, \nu ) \right] =  1 - \frac{2e^2}{\epsilon_0 k^2} \, \int \frac{\mathrm{d}^3q}{\left(2\pi\right)^3} \times \\ \times \frac{\left(f_{\vec{q} - \frac{\vec{k}}{2}} - f_{\vec{q} + \frac{\vec{k}}{2}} \right) \left( \hbar \tilde{\omega} - \left[ E_{\vec{q} + \frac{\vec{k}}{2}} - E_{\vec{q} - \frac{\vec{k}}{2}} \right] \right)}{\left( \hbar \tilde{\omega} - \left[ E_{\vec{q} + \frac{\vec{k}}{2}} - E_{\vec{q} - \frac{\vec{k}}{2}} \right] \right)^2 + \hbar^2 \nu_1^2} ,
\end{multline}

\begin{multline} 
  \label{eq_eps_rpa_im}
\operatorname{Im} \left[ \epsilon^\mathrm{RPA} (\vec{k}, \omega + i \, \nu ) \right] = \frac{2e^2}{\epsilon_0 k^2} \,  \int \frac{\mathrm{d}^3q}{\left(2\pi\right)^3} \, \times \\ \times \hbar \nu_1 \frac{ f_{\vec{q} - \frac{\vec{k}}{2}} - f_{\vec{q} + \frac{\vec{k}}{2}}  }{\left( \hbar \tilde{\omega} - \left[ E_{\vec{q} + \frac{\vec{k}}{2}} - E_{\vec{q} - \frac{\vec{k}}{2}} \right] \right)^2 + \hbar^2 \nu_1^2}.
\end{multline}
The shifted frequency $\tilde{\omega} = \omega - \nu_2$ is introduced here.

These integrals are performed across the entire momentum space and can therefore be shifted by an arbitrary vector $\vec{y}$ because for an integral of a function $G(\vec{x})$, which goes to $0$ as $|\vec{x}| \to \infty$, it holds that
\begin{equation} 
  \label{eq_int_trick}
\int_{\mathbb{R}^3} \mathrm{d}^3 x \, G(\vec{x}) = \int_{\mathbb{R}^3} \mathrm{d}^3 x \, G(\vec{x}- \vec{y}), \quad \mathrm{with} \, |\vec{y}| < \infty.
\end{equation}
Therefore, we can separate the integrand in Eqs.~\eqref{eq_eps_rpa_re} and~\eqref{eq_eps_rpa_im} into two summands with the Fermi occupation of the up- and down-shifted momentum, respectively. We further use Eq.~\eqref{eq_int_trick} to shift the momenta in the argument of the Fermi occupation to $\vec{q}$ in order to get $f_{\vec{q}}$ as a common prefactor for both summands. The momenta in the subscripts of the energy have to be shifted accordingly. This gives
\begin{multline} 
  \label{eq_eps_rpa_re2}
\operatorname{Re} \left[ \epsilon^\mathrm{RPA} (\vec{k}, \omega + i \, \nu ) \right] =  1 - \\ - \frac{2e^2}{\epsilon_0 k^2} 2\pi \int_0^\infty \frac{\mathrm{d}q}{(2\pi)^3} q^2 f_q \frac{m_e}{\hbar^2 k} \times \\ \times \int_{-1}^{1} \mathrm{d}z \bigg( \frac{\kappa - \frac{1}{2} \left( k + 2 q z  \right)}{\left( \kappa - \frac{1}{2} \left( k + 2 q z \right) \right)^2 + \Delta^2} - \\ - \frac{\kappa - \frac{1}{2} \left( -k + 2 q z  \right)}{\left( \kappa - \frac{1}{2} \left( -k + 2 q z \right) \right)^2 + \Delta^2} \bigg)
\end{multline}
for the real part and
\begin{multline} 
  \label{eq_eps_rpa_im2}
\operatorname{Im} \left[ \epsilon^\mathrm{RPA} (\vec{k}, \omega + i \, \nu ) \right] =   \\ 4 \pi \frac{e^2}{\epsilon_0 k^2} \int_0^\infty \frac{\mathrm{d}q}{(2\pi)^3} \frac{\nu_1 m_e^2}{\hbar^3 k^2} q^2 f_q \times \\ \times \int_{-1}^{1} \mathrm{d}z \bigg( \frac{1}{\left( \kappa - \frac{1}{2} \left( k + 2 q z \right) \right)^2 + \Delta^2} - \\ - \frac{1}{\left( \kappa - \frac{1}{2} \left( -k + 2 q z \right) \right)^2 + \Delta^2} \bigg)
\end{multline}
for the imaginary part.
Here, $\vec{k}$ was fixed in the $q_3$-direction and $z=\cos \theta$ where $\theta$ is the angle between $\vec{q}$ and $\vec{k}$. The shorthands $\kappa = \frac{\tilde{\omega} m_e}{\hbar k}$ and $\Delta = \frac{m_e \nu_1}{\hbar k}$ with the electron mass $m_e$ are introduced. The Fermi occupation can be pulled out of the angle integration as it only depends on the magnitude of the momentum.
The integral over the angle can be performed analytically in Eqs.~\eqref{eq_eps_rpa_re2} and~\eqref{eq_eps_rpa_im2}, giving
\begin{multline} 
  \label{eq_eps_rpa_re3}
\operatorname{Re} \left[ \epsilon^\mathrm{RPA} (\vec{k}, \omega + i \, \nu ) \right] =  1 + 2\pi \frac{m_e e^2}{\epsilon_0 \hbar^2 k^3}  \int_0^\infty \frac{\mathrm{d}q}{(2\pi)^3} q f_q \times \\ \times \ln \frac{\left(\Delta^2 + \left( \kappa - \frac{k}{2} - q \right)^2 \right) \left(\Delta^2 + \left( \kappa + \frac{k}{2} + q \right)^2 \right)}{\left(\Delta^2 + \left( \kappa - \frac{k}{2} + q \right)^2 \right) \left(\Delta^2 + \left( \kappa + \frac{k}{2} - q \right)^2 \right)}
\end{multline}
for the real part, and
\begin{multline} 
  \label{eq_eps_rpa_im3}
\operatorname{Im} \left[ \epsilon^\mathrm{RPA} (\vec{k}, \omega + i \, \nu ) \right] =  -4\pi \frac{m_e e^2}{\epsilon_0 \hbar^2 k^3}  \int_0^\infty \frac{\mathrm{d}q}{(2\pi)^3} q f_q \times \\ \times \bigg[ \arctan \left(\frac{\kappa - \frac{k}{2} - q}{\Delta} \right) + \arctan \left(\frac{\kappa + \frac{k}{2} + q}{\Delta} \right) - \\ - \arctan \left(\frac{\kappa - \frac{k}{2} + q}{\Delta} \right) - \arctan \left(\frac{\kappa + \frac{k}{2} - q}{\Delta} \right) \bigg]
\end{multline}
for the imaginary part of the RPA dielectric function modulated by a complex frequency. The remaining integration over $q$ has to be performed numerically.

\section{Expressions for the Born collision frequency}
\label{sec:app-born}

One of the most prominent approximations for the collision frequency is the Born collision frequency~\cite{Reinholz2000}
\begin{multline} 
  \label{eq_born_coll}
\nu^{\mathrm{Born}} (\omega ) =  - i \frac{\epsilon_0 n_\mathrm{i} \Omega^2}{6 \pi^2 e^2 n_e m_e}   \int_0^\infty \mathrm{d}q \, q^6 \times \\ \times V_{e\mathrm{i}}^2(q) S_\mathrm{ii}(q) \frac{1}{\omega} \left[ \epsilon^{\mathrm{RPA}}(q, \omega) - \epsilon^{\mathrm{RPA}}(q, 0) \right],
\end{multline}
with the ion density $n_\mathrm{i}$, the electron density $n_e$ and the normalization volume $\Omega$. There are different approximations for the electron-ion potential $V_{e\mathrm{i}}$ and the static structure factor $S_\mathrm{ii}$. The potential can be approximated by the screened Coulomb potential with the Debye-Hückel or Thomas-Fermi screening parameter depending on the density and temperature regime considered.
Approaches to the structure factor range from the assumption of a homogeneous electron gas ($S_\mathrm{ii}(q) = 1$) or analytic models like the Debye-Hückel theory to more sophisticated methods like the hypernetted-chain (HNC) equation or MD simulations. Here, we use the potential
\begin{equation}
V_{e\mathrm{i}}(q) = \frac{V^\mathrm{Coulomb}_{e\mathrm{i}}(q)}{\epsilon^\mathrm{RPA} (q, 0)} = - \frac{e_e e_\mathrm{i}}{\epsilon_0 \Omega} \frac{1}{q^2 \, \epsilon^\mathrm{RPA}(q,0)}
\end{equation}
and the static structure factor we calculate from our DFT-MD simulations.
Equation~\eqref{eq_born_coll} is computed by directly calculating its real part and subsequently performing the Kramers-Kronig~\cite{Kronig1926, Kramers1927} transformation to arrive at the imaginary part.

\section{Temperature determination via Laplace transform}
\label{sec:app-laplace}

\begin{figure}[t]
\center{\includegraphics[angle=0,width=1.0\linewidth]{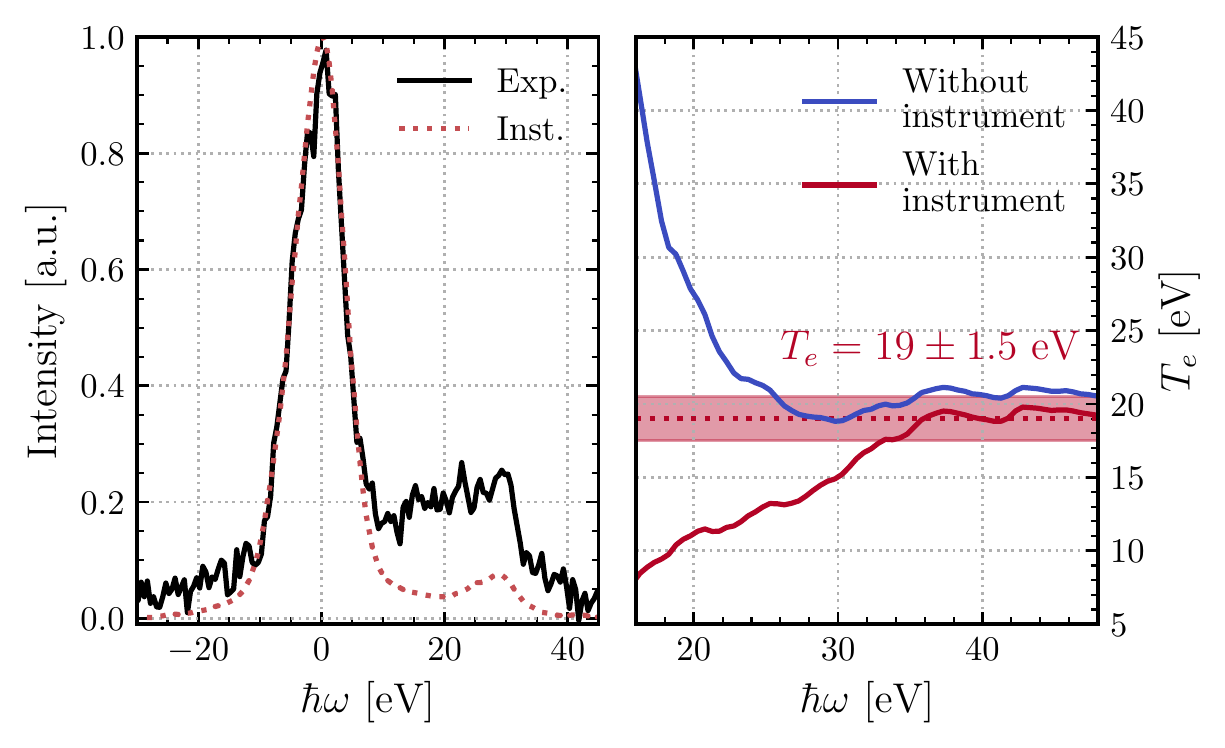}}
\caption{The left panel shows the scattering intensity and the instrument function from Ref.~\onlinecite{Doeppner2009}. The right panel shows the inferred electron temperature accoring to Ref.~\onlinecite{Dornheim2022_natcom}.}
\label{fig:DoeppnerLaplace}
\end{figure}

We employ the recently proposed temperature diagnostic based on a two-sided Laplace transform~\cite{Dornheim2022_natcom} to inferr the temperature from experiment performed by Döppner~\textit{et~al.} The left panel of Fig.~\ref{fig:DoeppnerLaplace} shows the scattering data and the instrument function, while the right panel shows the inferred temperature according to the procedure described in Ref.~\onlinecite{Dornheim2022_natcom}. The x axis denotes the energy up to which the two-sided Laplace transform is performed. A convergence is observed beyond 40~eV and the electron temperature is determined to be $19 \pm 1.5$~eV. This values agrees within errorbars with the electron temperature of 18~eV, originally determined by Döppner~\textit{et~al.} We, therefore, exclude the electron temperature from the fitting procedure for this experiment.

\bibliography{main.bib}

\end{document}